\documentclass[preprint,aps,amsmath,showpacs,showkeys,nofootinbib, 
superscriptaddress]{revtex4} 
\usepackage{epsfig}
\usepackage{graphicx}
\begin{document} 
\title{\textbf{Effect of thresholds on the width of three-body resonances}}
\author{H. Garcilazo} 
\email{humberto@esfm.ipn.mx} 
\affiliation{Escuela Superior de F\' \i sica y Matem\'aticas, \\ 
Instituto Polit\'ecnico Nacional, Edificio 9, 
07738 M\'exico D.F., Mexico} 
\author{A.~Valcarce} 
\email{valcarce@usal.es} 
\affiliation{Departamento de F\'\i sica Fundamental and IUFFyM,\\ 
Universidad de Salamanca, E-37008 Salamanca, Spain}
\date{\today} 
\begin{abstract}
It has been recently reported an intriguing theoretical result
of a narrow three-body resonance with a large 
available phase space~\cite{GAR4}.
The resonance was reported in the $N\Lambda\Lambda-\Xi NN$ system
near the $\Xi d$ threshold, having a 
very small width in spite of 
the open $N\Lambda\Lambda$ channel lying around 23 MeV 
below the $\Xi NN$ channel. We use first-order perturbation theory as a
plausible argument to explain this behavior. We apply
our result to realistic local interactions.
Other systems involving several thresholds are likely to 
follow the same behavior. 
\end{abstract}
\maketitle 

\section{Introduction}
The coupled $N\Lambda\Lambda-\Xi NN$ system in the dominant
S-wave configuration has the quantum numbers 
$(I,J^P)=(\frac{1}{2},\frac{1}{2}^+)$, since the coupling 
between the lower ($N\Lambda\Lambda$) and upper
($\Xi NN$) components of the system is via the 
$\Lambda\Lambda - \Xi N$ two-body channel with quantum numbers
$(i,j^p)=(0,0^+)$. Therefore, if one adds an additional 
nucleon also in S-wave, the three-body system will have the
quantum numbers $(I,J^P)=(\frac{1}{2},\frac{1}{2}^+)$. 

The possible existence of a stable bound state of the coupled
three-body system $N\Lambda\Lambda-\Xi NN$ was first studied
in Refs.~\cite{GAR1,GAR2,GAR3} using the two-body
interactions derived in a constituent quark model framework~\cite{VAL1}. 
In that model, the coupled two-body
subsystem $\Lambda\Lambda-\Xi N$ in the $(i,j^p)=(0,0^+)$ 
channel (the $H$ dibaryon channel) is bound with a binding energy
of 6.4 MeV. Thus, in order to search for bound-state solutions
of the three-body equations one just needs to calculate the 
(real) Fredholm determinant for energies below the $HN$ threshold which indeed 
leds to a stable bound-state solution at about 0.5 MeV below that
threshold~\cite{GAR1}.

However, the most recent analysis of the 
quark mass dependence of the $H$ dibaryon in 
$\Lambda\Lambda$ scattering~\cite{Sha11,Yam16}
point to the $H$ dibaryon being a resonance above the $\Lambda\Lambda$
threshold. Thus, in order to study possible bound or resonant
states of the $N\Lambda\Lambda-\Xi NN$ system one must solve
the three-body equations in the complex plane, which makes the 
numerical problem much harder to deal with. This was done
in Ref.~\cite{GAR4} using simple separable potentials fitted
to the low-energy data of the most recent update of the 
ESC08 Nijmegen potential~\cite{RIJ1,RIJ2}, that give account 
of the pivotal results of strangeness $-2$ physics, the NAGARA~\cite{Tak01} 
and the KISO~\cite{Naa15} events. 

\section{Formalism}
The results of Ref.~\cite{GAR4} were obtained  taking the nucleon mass as the
average of the proton and neutron mass, and the $\Xi$ mass as the average 
of $\Xi^0$ and $\Xi^-$ mass. Thus, the $\Xi NN$ and $\Xi d$ thresholds are
25.604 MeV and 23.420 MeV above the $N\Lambda\Lambda$ threshold,
respectively. 

\subsection{The problem}
It was found in Ref.~\cite{GAR4} that
the three-body resonance lies at $E_0=23.408-i\, 0.045$ MeV 
measured with respect to the $N\Lambda\Lambda$ threshold,
which is 0.012 MeV below the $\Xi d$ threshold. Thus,
it is a $N\Lambda\Lambda$ resonance 
as seen from the lower component or a $\Xi NN$
quasibound state as seen from the upper component. The most
interesting feature of this result is the very small width. 
If one neglects the $\Lambda\Lambda-\Xi N$ 
$(0,0^+)$ channel, which is responsible for the coupling
between the $N\Lambda\Lambda$ and $\Xi NN$ components, the resonance
becomes a bound state of the $\Xi NN$ system at $E_0=23.386$ MeV
with $E_0$ measured with respect to the $N\Lambda\Lambda$ threshold.
Thus, the effect of the $\Lambda\Lambda-\Xi N$ $(0,0^+)$ channel
is to change the three-body eigenvalue by 
\begin{equation}
\delta E=0.022-i\, 0.045\,\,\,\, {\rm MeV}, 
\label{eq0}
\end{equation}
indicating that the lower three-body channel effectively 
acts as a perturbation. This is somewhat intriguing since the
$\Lambda\Lambda-\Xi N$ $(0,0^+)$ interaction is not small 
(see Tables II and  III of Ref.~\cite{GAR4}). 

\subsection{The proposed explanation}
In order to provide a plausible argument to explain 
the above result, we will consider first-order perturbation theory 
taking the $\Xi NN$ channel as the main interaction and
the contribution of the lower channels $N\Lambda\Lambda$ as the perturbation.
The argument is very simple. The small 
binding energy of the $\Xi NN$ system causes the unperturbed 
wave function to have a very long range in coordinate space
while the perturbation is a short-range operator so that the
overlap between them is quite small, which results in $\delta E$ 
being small.
Thus, we will calculate
\begin{equation}
\delta E = \frac{\langle\Psi_0\mid\delta V\mid \Psi_0\rangle} 
                {\langle\Psi_0\mid\Psi_0\rangle},
\label{eq1}
\end{equation}
where $\mid\Psi_0\rangle$ is the (real) $\Xi NN$ wave function 
\begin{equation}
\mid\Psi_0\rangle=G_0^{\Xi NN}\left(\mid U_{1;\Xi}^{NN}\rangle
+\mid U_{2;N}^{\Xi N}\rangle+\mid U_{3;N}^{\Xi N}\rangle\right),
\label{eq2}
\end{equation}
with $G_0^{\Xi NN}$ the Green's function for three free particles,
$\mid U_{1;\Xi}^{NN}\rangle$ is the Faddeev component where the two
nucleons interact last with the $\Xi$ as spectator, and similarly the 
other two Faddeev components. They are determined by the last two 
Eqs. (14) of Ref.~\cite{GAR4} neglecting altogether
the $(i,j^p)=(0,0^+)$ channel.

\begin{figure*}[t]
\vspace*{-2cm}
\resizebox{14.cm}{17.cm}{\includegraphics{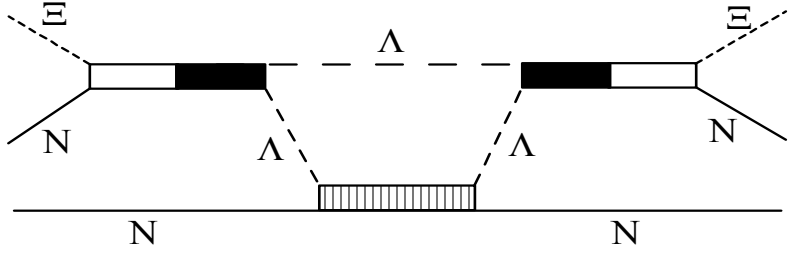}}
\vspace*{-10.5cm}
\caption{Complex perturbation $\delta V$ given in Eq.~(\ref{eq3}).}
\label{fig1}
\end{figure*}

$\delta V$ is the (complex) perturbation given in lowest order by,
\begin{equation}
\delta V = \sum_{j=2}^3
t_{i1;N}^{N\Xi - \Lambda\Lambda}G_0^{N\Lambda\Lambda}
t_{j;\Lambda}^{N \Lambda}
G_0^{N\Lambda\Lambda}t_{1i;N}^{\Lambda\Lambda - N\Xi};\,\,\,\,\,i=2,3,
\label{eq3}
\end{equation}
which is shown graphically in Fig.~\ref{fig1}. We have used the 
convention that in both three-body sectors the two identical 
particles are labeled 2 and 3 with particle 1 being the different one. 
Using Eqs.~(\ref{eq1}-\ref{eq3}) and taking into account the identity 
of particles 2 and 3 one obtains, 
\begin{eqnarray}
\langle\Psi_0\mid\delta V\mid\Psi_0\rangle &=& 4\left( 
 \langle U_{1;\Xi}^{NN}\mid G_0^{\Xi NN}
 t_{31;N}^{N\Xi - \Lambda\Lambda}G_0^{N\Lambda\Lambda}+
 \langle U_{3;N}^{\Xi N}\mid G_0^{\Xi NN}
 t_{21;N}^{N\Xi - \Lambda\Lambda}G_0^{N\Lambda\Lambda}\right)
 \nonumber \\ &\times& t_{3;\Lambda}^{N \Lambda}\left(
 G_0^{N\Lambda\Lambda}t_{13;N}^{\Lambda\Lambda - N\Xi}G_0^{\Xi NN}
 \mid U_{1;\Xi}^{NN}\rangle +             
 G_0^{N\Lambda\Lambda}t_{12;N}^{\Lambda\Lambda - N\Xi}G_0^{\Xi NN}
 \mid U_{3;N}^{\Xi N}\rangle\right). 
\label{eq4}
\end{eqnarray}
In Eq. (\ref{eq4}), terms of the form $\langle U_{3;N}^{\Xi N}\mid G_0^{\Xi NN}
t_{31;N}^{N\Xi - \Lambda\Lambda}$ and $t_{13;N}^{\Lambda\Lambda - N\Xi}G_0^{\Xi NN}
 \mid U_{3;N}^{\Xi N}\rangle$ do not contribute due to
the orthogonality of the spin-isospin states $_\alpha\langle (12)3\mid (12)3\rangle_\beta=\delta_{\alpha\beta}$
since the amplitude $t_{31;N}^{N\Xi - \Lambda\Lambda}$ which belongs
to the perturbation corresponds to the two-body channel
$\beta=(0,0^+)$ while the component $\langle U_{3;N}^{\Xi N}\mid$ of the unperturbed wave function
involves only the two-body channels $\alpha\ne (0,0^+)$.

The Green's function that appears in the perturbation term~(\ref{eq3})
is given explicitly by,
\begin{equation}
G_0^{N\Lambda\Lambda}(p_iq_i)=\frac{1}{E-p_i^2/2\eta_i-q_i^2/2\nu_i
+i\epsilon},
\label{eq5}
\end{equation}
where $p_i$ and $q_i$ are the Jacobi momenta and $\eta_i$ and $\nu_i$
the corresponding reduced masses of the various configurations.
Since $E$ is a positive number this function is singular and 
moreover it has an imaginary part.
The Green's function attached to the Faddeev components of
the unperturbed wave function~(\ref{eq2}), on the other hand, is given by
\begin{equation}
G_0^{\Xi NN}(p_iq_i)=\frac{1}{E+\Delta E-p_i^2/2\eta_i-q_i^2/2\nu_i},
\label{eq6}
\end{equation}
with
\begin{equation}
\Delta E = m_\Lambda-m_\Xi,
\label{eq7}
\end{equation}
so that $E+\Delta E$ is a small negative number and the function 
in Eq.~(\ref{eq6}) is real and it has no singularity 
although it is sharply peaked at low momenta. In addition, 
the Faddeev amplitudes $U_{1;\Xi}^{NN}$ and $U_{3;N}^{\Xi N}$
are also peaked at low momenta in the $NN$
$(0,1^+)$ and $\Xi N$ $(1,1^+)$ channels,
corresponding to the deuteron and $D^*$ bound states, respectively, which lie
very close to threshold. Thus, as mentioned above,
the unperturbed wave function has a long range in coordinate space while the
perturbation term has the short-range characteristic of hadronic systems.
Consequently, the overlap between both terms in Eq.~(\ref{eq4}) is very
small, rendering $\delta E$ small.
\begin{figure}[b]
\vspace*{-1.5cm}
\resizebox{7.5cm}{11.5cm}{\includegraphics{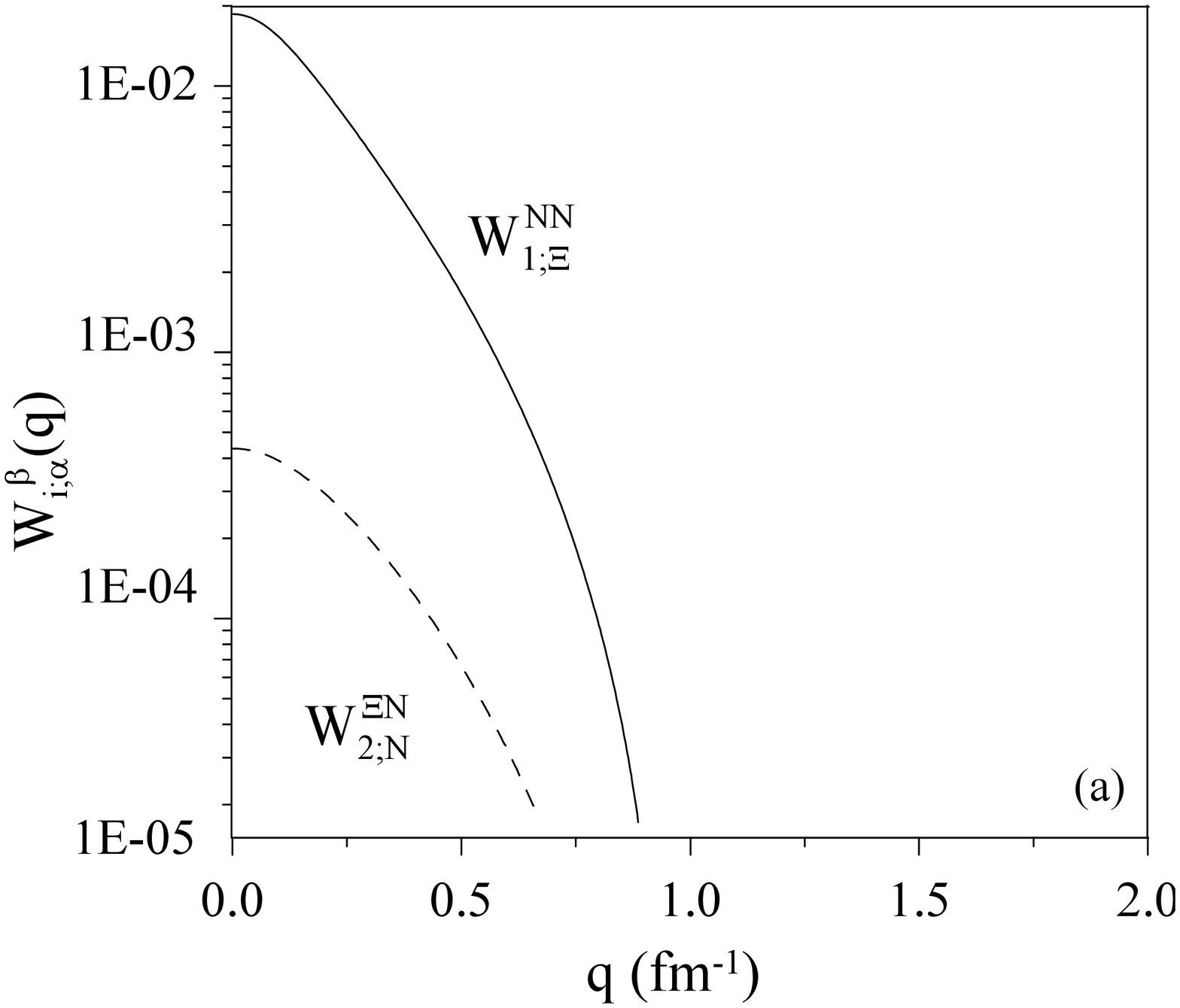}}\hspace{0.75cm}
\resizebox{7.5cm}{11.5cm}{\includegraphics{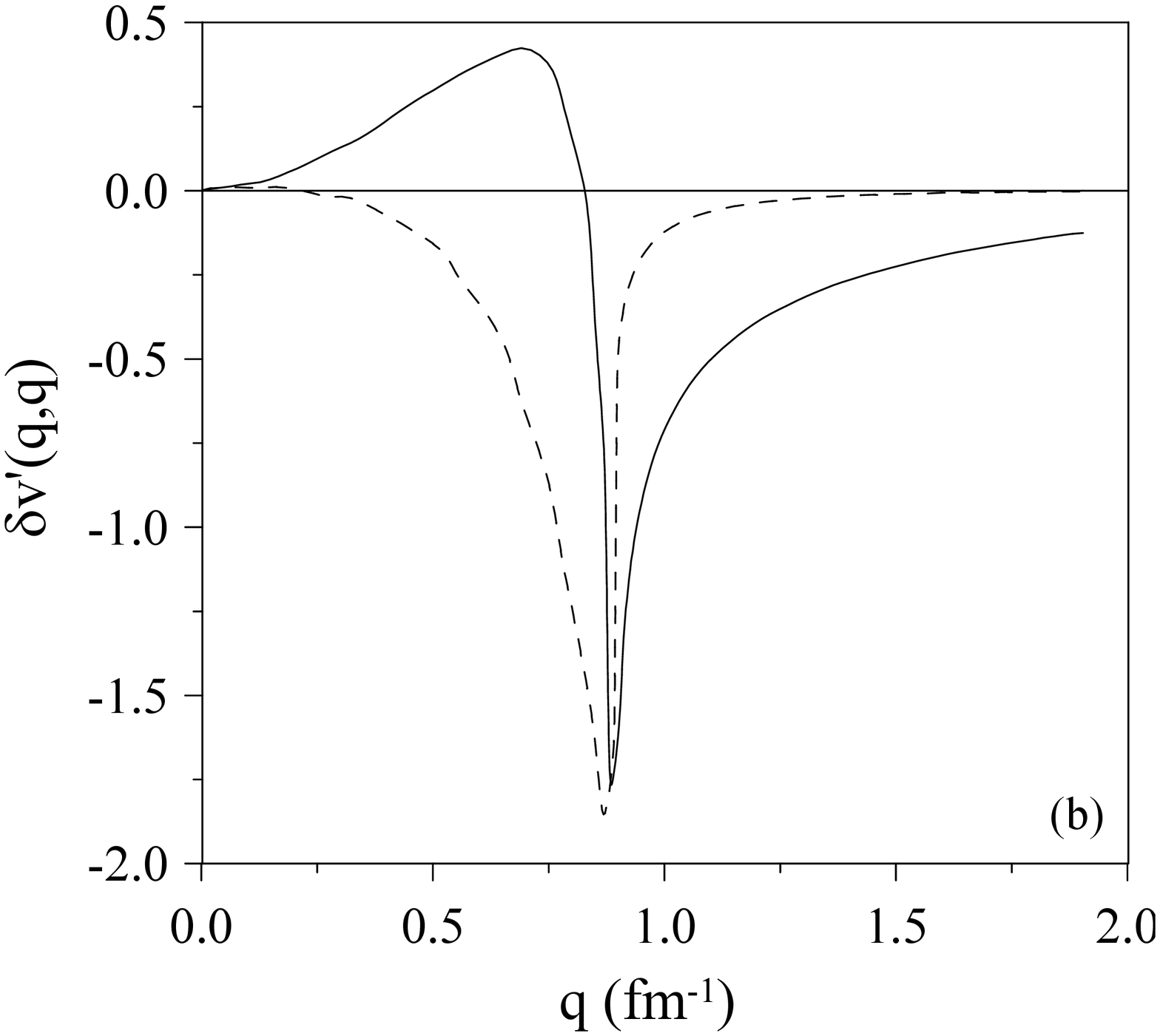}}
\vspace*{-2.5cm}
\caption{(a) Wave functions $W_{1;\Xi}^{NN}(q)$ and $W_{2;N}^{\Xi N}(q)$, in arbitrary units,
for the two dominant channels: $NN$ $(0,1^+)$ (solid line) and $N\Xi$ $(1,1^+)$ (dashed line). 
(b) Real (solid line) and imaginary (dashed line) parts of the 
diagonal perturbation term $\delta v^{\prime}(q,q)$, in arbitrary units.} 
\label{fignew}
\end{figure}

In order to show explicitly this behavior let us consider one of the terms of Eq.~(\ref{eq4}),
\begin{equation}
\delta v=
 \langle U_{1;\Xi}^{NN}\mid G_0^{\Xi NN}
 t_{31;N}^{N\Xi - \Lambda\Lambda}G_0^{N\Lambda\Lambda}
 t_{3;\Lambda}^{N \Lambda}
 G_0^{N\Lambda\Lambda}t_{12;N}^{\Lambda\Lambda - N\Xi}G_0^{\Xi NN}
 \mid U_{3;N}^{\Xi N}\rangle \, .
\label{eqx1}
\end{equation}
Since in the separable model of Ref.~\cite{GAR4} one has that,
\begin{eqnarray}
t_{31;N}^{N\Xi - \Lambda\Lambda} & = & g_3^{N\Xi}
\tau_{31;N}^{N\Xi - \Lambda\Lambda} g_1^{\Lambda\Lambda} \, , \nonumber \\
t_{12;N}^{\Lambda\Lambda - N\Xi} & = & g_1^{\Lambda\Lambda}
\tau_{12;N}^{\Lambda\Lambda - N\Xi} g_2^{N\Xi} \, ,
\label{eqx2}
\end{eqnarray}
one can rewrite Eq.~(\ref{eqx1}) as,
\begin{equation}
\delta v=
 \langle W_{1;\Xi}^{NN}\mid \delta v^{\prime}
 \mid W_{2;N}^{\Xi N}\rangle \, ,
\label{eqx3}
\end{equation}
where
\begin{align}
 \langle W_{1;\Xi}^{NN}\mid  = & 
 \langle U_{1;\Xi}^{NN} G_0^{\Xi NN}g_3^{N\Xi}\mid \, , \nonumber \\
 \mid W_{2;N}^{\Xi N}\rangle  = &  \mid g_2^{N\Xi}
 G_0^{\Xi NN} U_{3;N}^{\Xi N}\rangle \, , \label{eqx6} \\
\delta v^{\prime}   = 
\tau_{31;N}^{N\Xi - \Lambda\Lambda} 
 g_1^{\Lambda\Lambda}
&G_0^{N\Lambda\Lambda}
 t_{3;\Lambda}^{N \Lambda}
 G_0^{N\Lambda\Lambda}
 g_1^{\Lambda\Lambda}
\tau_{12;N}^{\Lambda\Lambda - N\Xi} \nonumber \, .
\end{align}
The expressions~(\ref{eqx6}) depend only on
the variables $q_i$, i.e.,
\begin{eqnarray}
  W_{1;\Xi}^{NN}(q_3)  & \equiv &
 \langle W_{1;\Xi}^{NN}\mid q_3\rangle \, , \nonumber \\
 W_{2;N}^{\Xi N}(q_2)  & \equiv &
\langle q_2 \mid W_{2;N}^{\Xi N}\rangle \, , \\
\delta v^{\prime}(q_3,q_2) & \equiv &
 \langle q_3\mid \delta v^{\prime} \mid q_2\rangle \, . \nonumber
\label{eqx9}
\end{eqnarray}
We show in Fig.~\ref{fignew}(a) the wave functions
$W_{1;\Xi}^{NN}(q)$ and  $W_{2;N}^{\Xi N}(q)$ for the two dominant
channels $NN$ $(0,1^+)$ and $N\Xi$ $(1,1^+)$, respectively,
and in Fig.~\ref{fignew}(b) the diagonal perturbation term
$\delta v^{\prime}(q,q)$, where one can see clearly this behavior (note
the logarithmic scale of the wave function).

\section{Results}
\subsection{The separable model}
If we now apply the formalism of the previous section to the separable
potential model of Ref.~\cite{GAR4}, it gives
\begin{equation}
\delta E=0.014-i \, 0.015\,\,\,\, {\rm MeV}, 
\label{eq9}
\end{equation}
which is of the same order of magnitude as the result of the exact
calculation given by Eq.~(\ref{eq0}). This shows that the small
value of $\delta E$, and consequently the very small width,
can be understood as resulting from the fact that
the $N\Lambda\Lambda$ channel acts effectively as a perturbation
to the $\Xi NN$ channel when the resonance lies very near the
$\Xi NN$ threshold. 

\subsection{Application to the local model}
\begin{figure}[t]
\vspace*{-0.5cm}
\mbox{\epsfxsize=100mm\epsffile{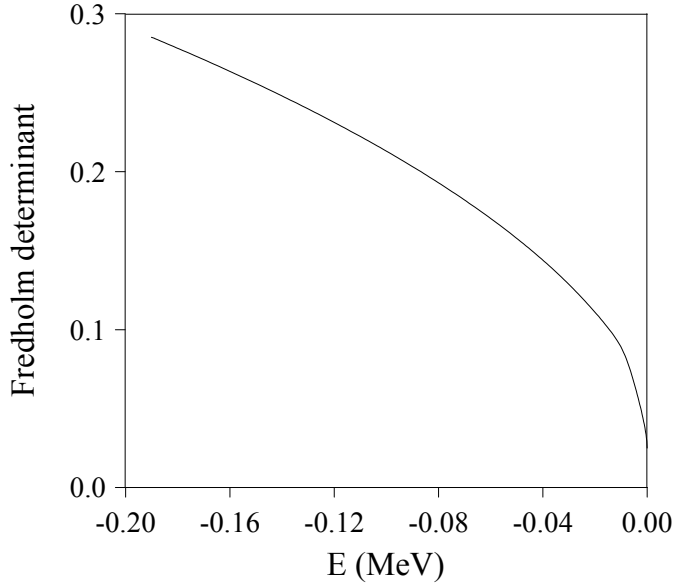}}
\vspace*{-6cm}
\caption{Fredholm determinant of the $\Xi NN$
system for energies very near the $\Xi d$ threshold.}
\label{fig2}
\end{figure}

The behavior of three-body resonances lying very near 
the upper channel threshold described in the previous section
does not depend on the interactions being separable,
but it is completely general. Thus, we have also used the Malfliet-Tjon 
type local potentials~\cite{MALF} of the $NN$ 
subsystem constructed in Ref.~\cite{FRIA} and those of the $\Xi N$
subsystem constructed in Ref.~\cite{GAR5}, based in the most 
recent update of the Nijmegen ESCO8 potentials~\cite{RIJ2}. 
We show the Fredholm determinant of the $\Xi NN$
system in Fig.~\ref{fig2} for energies very near 
the $\Xi d$ threshold where as one can see 
the $\Xi NN$ state lies less than  0.01 MeV above the $\Xi d$ threshold, so
that both the separable and local models predict the resonance very
near the $\Xi d$ threshold and consequently, as we have just shown, 
will have a very small width.

\section{Outlook}

In this letter we have presented a plausible argument to explain
the small width of a three-body resonance in a coupled two-channel system
lying close to the upper channel in spite of being open the lower one. 
This is an intriguing result,
since the available phase space of the decay channel is quite
large, around 23 MeV. We use first-order perturbation theory 
to explain this behavior. We have applied our result to realistic 
local interactions.
 
Let us finally comment that the mechanism we have discussed in this work
could also help in understanding the narrow width of some experimental 
resonances found in the heavy hadron spectra, whose assumed internal structure 
allow them to split into several different channels~\cite{Che16,Pil17}. It has been 
explained in Ref.~\cite{Car12} how systems with an internal 
structure $Q\bar Q n\bar n$, where $n$ stands for a light quark, could 
either split into $(Q \bar n) - (n \bar Q)$ or
$(Q \bar Q) - (n \bar n)$. For $Q=c$ or $Q=b$ the $(Q \bar Q) - (n \bar n)$ 
threshold is lower than the $(Q \bar n) - (n \bar Q)$, the mass difference
augmenting when increasing the mass of the heavy quark.
Such experimental behavior
can be simply understood within the constituent quark model
with a Cornell-like potential~\cite{Car12,Clo03}. 
Thus, the possibility of finding meson-antimeson molecules, $(Q \bar n) - (n \bar Q)$, contributing to 
the heavy meson spectra becomes more and more difficult when increasing the mass
of the heavy flavor, due to the lowering of the mass of the $(Q \bar Q) - (n\bar n)$ 
threshold. This would make the system dissociate immediately. In such cases,
the presence of attractive meson-antimeson threshold together with the 
arguments we have drawn in this work, hint to a possible explanation of a narrow 
width of some of the $XYZ$ states lying close to the $(Q \bar n) - (n \bar Q)$ threshold 
as a meson-antimeson molecule. Similar arguments could be handled for 
the LHCb pentaquarks, what requires a careful analysis
in the models used for the study of these states.

\acknowledgments 
This work has been partially funded by COFAA-IPN (M\'exico), 
by the Spanish Ministerio de Econom{\'\i}a, Industria y Competitividad
and EU FEDER under Contracts No. FPA2013-47443-C2-2-P, FPA2015-69714-REDT 
and FPA2016-77177-C2-2-P, and by Junta de Castilla y Le\'on under 
Contract No. SA041U16.

\end{document}